\documentclass{ws-ijmpd}

\newcommand{\z}[0]{{\text z}}
\newcommand{\ta}[0]{{t_\textit{O}}}
\newcommand{\bp}[0]{{\beta_\textit{O}}}

\newcommand{\lp}[0]{{\lambda_\textit{O}}}
\newcommand{\ie}[0]{{i.e.}, }

\newcommand{\figref}[1]{{Fig.~\ref{#1}}}
\newcommand{\Figref}[1]{{Fig.~\ref{#1}}}
\newcommand{\eqnref}[1]{{Eq.~\eqref{#1}}}
\newcommand{\Eqnref}[1]{{Eq.~\eqref{#1}}}
\newcommand{\eqnsref}[2]{{Eqs.~\eqref{#1} and \eqref{#2}}}
\def\phi{\varphi}
\usepackage{graphics}
\newcommand{\image}[4]
{
\begin{figure}[htb]
\centering
\makebox{
  \resizebox
      {#2cm}{!}
     {\includegraphics{#1}}
}
\caption{#4}
\label{#3}
\end{figure}
}
\def\astrobj#1{#1}

\begin{document}

\markboth{Manoj Thulasidas}
{DRAGNs and GRBs Luminal Booms?}

%
\catchline{}{}{}{}{}
%

\title{ARE RADIO SOURCES AND GAMMA RAY BURSTS LUMINAL BOOMS?}

\author{MANOJ THULASIDAS}

\address{Neural Signal Processing Lab, Institute for Infocomm
  Research,\\
  National University of Singapore,\\
  21 Heng Mui Keng Terrace, Singapore 119613\\
  manoj@thulasidas.com}

\maketitle

\begin{history}
\received{Day Month Year}
\revised{Day Month Year}
\comby{Managing Editor}
\end{history}

\begin{abstract}
  The softening of a Gamma Ray Burst (GRB) afterglow bears remarkable
  similarities to the frequency evolution in a sonic boom. At the
  front end of the sonic boom cone, the frequency is infinite, much
  like a GRB.  Inside the cone, the frequency rapidly decreases to
  infrasonic ranges and the sound source appears at two places at the
  same time, mimicking the double-lobed radio sources.  Although a
  ``luminal'' boom violates the Lorentz invariance and is therefore
  forbidden, it is tempting to work out the details and compare them
  with existing data.  This temptation is further enhanced by the
  observed superluminality in the celestial objects associated with
  radio sources and some GRBs.  In this article, we calculate the
  temporal and spatial variation of observed frequencies from a
  hypothetical luminal boom and show remarkable similarity between our
  calculations and current observations.
\end{abstract}

\keywords{special relativity;
light travel time effect;
gamma rays bursts;
radio sources.}

\section{Introduction}\label{S1}

A sonic boom is created when an object emitting sound passes through
the medium faster than the speed of sound in that medium. As the
object traverses the medium, the sound it emits creates a conical
wavefront, as shown in \figref{Supersonic}.  The sound frequency at
this wavefront is infinite because of the Doppler shift.  The
frequency behind the conical wavefront drops dramatically and soon
reaches the infrasonic range.  This frequency evolution is remarkably
similar to afterglow evolution of a gamma ray burst (GRB).

\image{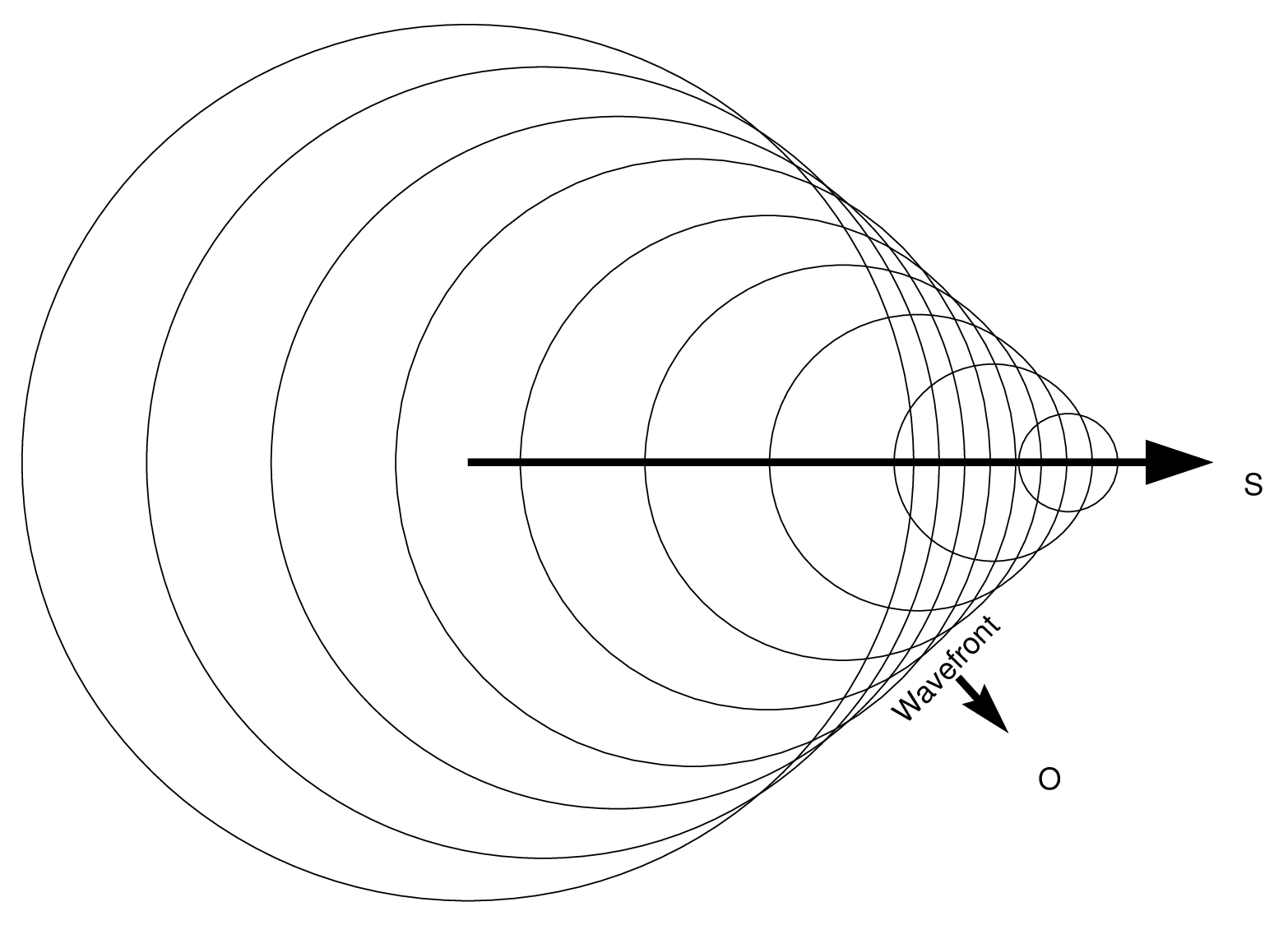}{8}{Supersonic}{The frequency evolution of sound
  waves as a result of the Doppler effect in supersonic motion.  The
  supersonic object S is moving along the arrow.  The sound waves are
  ``inverted'' due to the motion, so that the waves emitted at two
  different points in the trajectory merge and reach the observer (at
  O) at the same time.  When the wavefront hits the observer, the
  frequency is infinity. After that, the frequency rapidly decreases.}

Gamma Ray Bursts are very brief, but intense flashes of $\gamma$ rays
in the sky, lasting from a few milliseconds to several
minutes,\cite{GRB1} and are currently believed to emanate from
cataclysmic stellar collapses.  The short flashes (the prompt
emissions) are followed by an afterglow of progressively softer
energies. Thus, the initial $\gamma$ rays are promptly replaced by
X-rays, light and even radio frequency waves. This softening of the
spectrum has been known for quite some time,\cite{GRB5} and was first
described using a hypernova (fireball) model.  In this model, a
relativistically expanding fireball produces the $\gamma$ emission,
and the spectrum softens as the fireball cools down.\cite{piran2}  The
model calculates the energy released in the $\gamma$ region as
$10^{53}$--$10^{54}$ ergs in a few seconds.  This energy output is
similar to about 1000 times the total energy released by the sun over
its entire lifetime.

More recently, an inverse decay of the peak energy with varying time
constant has been used to empirically fit the observed time evolution
of the peak energy\cite{GRB3,GRB4} using a collapsar model.  According
to this model, GRBs are produced when the energy of highly
relativistic flows in stellar collapses are dissipated, with the
resulting radiation jets angled properly with respect to our line of
sight.  The collapsar model estimates a lower energy output because
the energy release is not isotropic, but concentrated along the jets.
However, the rate of the collapsar events has to be corrected for the
fraction of the solid angle within which the radiation jets can appear
as GRBs.  GRBs are observed roughly at the rate of once a day.  Thus,
the expected rate of the cataclysmic events powering the GRBs is of
the order of $10^4$--$10^6$ per day. Because of this inverse
relationship between the rate and the estimated energy output, the
total energy released per observed GRB remains the same.

If we think of a GRB as an effect similar to the sonic boom in
supersonic motion, the assumed cataclysmic energy requirement becomes
superfluous.  Another feature of our perception of supersonic object
is that we hear the sound source at two different location as the same
time, as illustrated in \figref{dragn2-0}. This curious effect takes
place because the sound waves emitted at two different points in the
trajectory of the supersonic object reach the observer at the same
instant in time.  The end result of this effect is the perception of a
symmetrically receding pair of sound sources, which, in the luminal
world, is a good description of symmetric radio sources (Double
Radio source Associated with Galactic Nucleus or DRAGN).

\image{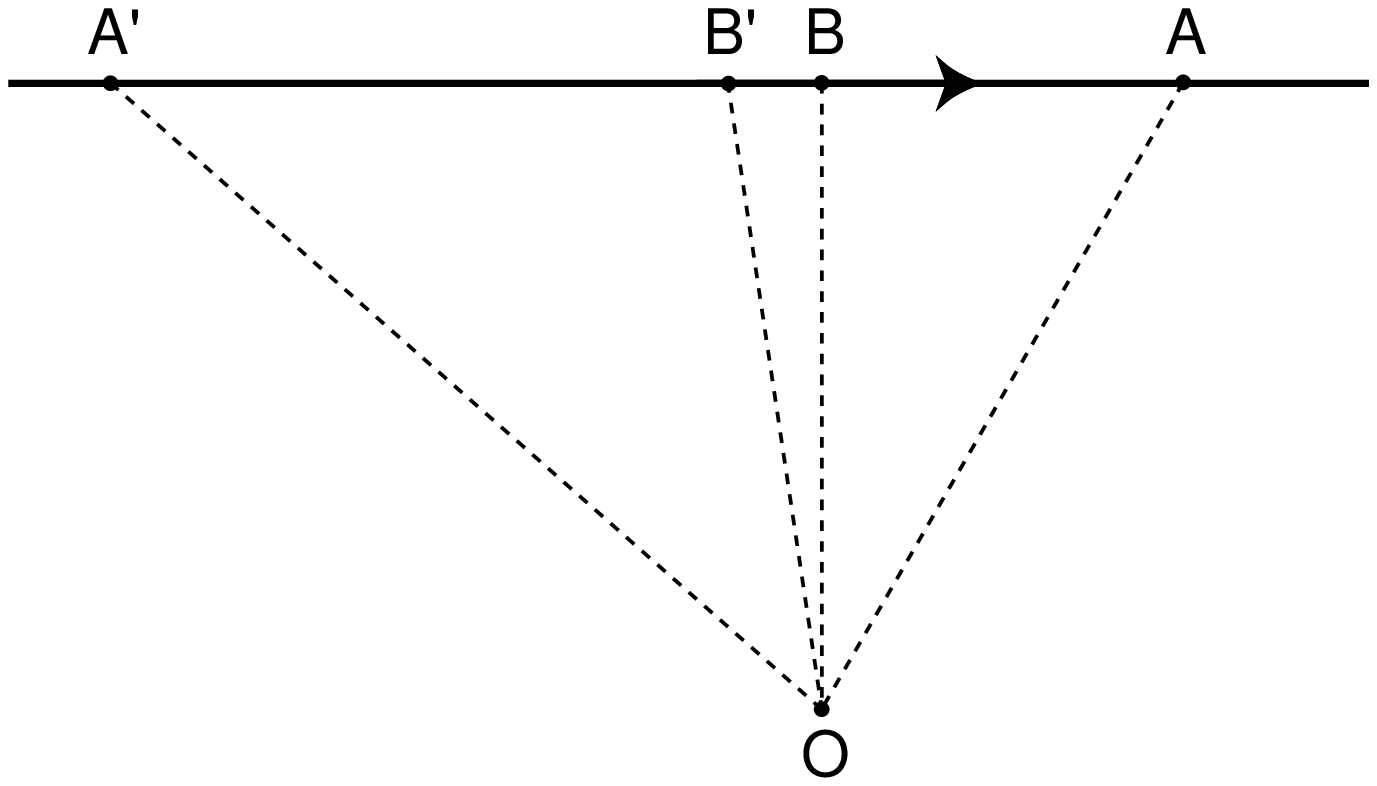}{8}{dragn2-0}{The object is flying from $A'$ to $A$
  through $B'$ and $B$ at a constant supersonic speed. Imagine that
  the object emits sound during its travel. The sound emitted at the
  point $B'$ (which is near the point of closest approach $B$) reaches
  the observer at $O$ before the sound emitted earlier at $A'$.  The
  instant when the sound at an earlier point $A'$ reaches the
  observer, the sound emitted at a much later point $A$ also reaches
  $O$.  So, the sound emitted at $A$ and $A'$ reaches the observer at
  the same time, giving the impression that the object is at these two
  points at the same time.  In other words, the observer hears two
  objects moving away from $B'$ rather than one real object.}

Radio Sources are typically symmetric and seem associated with
galactic cores, currently considered manifestations of space-time
singularities or neutron stars.  Different classes of such objects
associated with Active Galactic Nuclei (AGN) were found in the last
fifty years.  \Figref{cyga} shows the radio galaxy Cygnus
A,\cite{cyga} an example of such a radio source and one of the
brightest radio objects.  Many of its features are common to most
extragalactic radio sources: the symmetric double lobes, an indication
of a core, an appearance of jets feeding the lobes and the hotspots.
Refs.~\refcite{hotspot1} and \refcite{hotspot2} have reported more
detailed kinematical features, such as the proper motion of the
hotspots in the lobes.

Symmetric radio sources (galactic or extragalactic) and GRBs may
appear to be completely distinct phenomena.  However, their cores show
a similar time evolution in the peak energy, but with vastly different
time constants.  The spectra of GRBs rapidly evolve from $\gamma$
region to an optical or even RF afterglow, similar to the spectral
evolution of the hotspots of a radio source as they move from the core
to the lobes.  Other similarities have begun to attract attention in
the recent years.\cite{GRB7}

This article explores the similarities between a hypothetical
``luminal'' boom and these two astrophysical phenomena, although such
a luminal boom is forbidden by the Lorentz invariance. Treating GRB as
a manifestation of a hypothetical luminal boom results in a model that
unifies these two phenomena and makes detailed predictions of their
kinematics.

\section{Symmetric Radio Sources}

\image{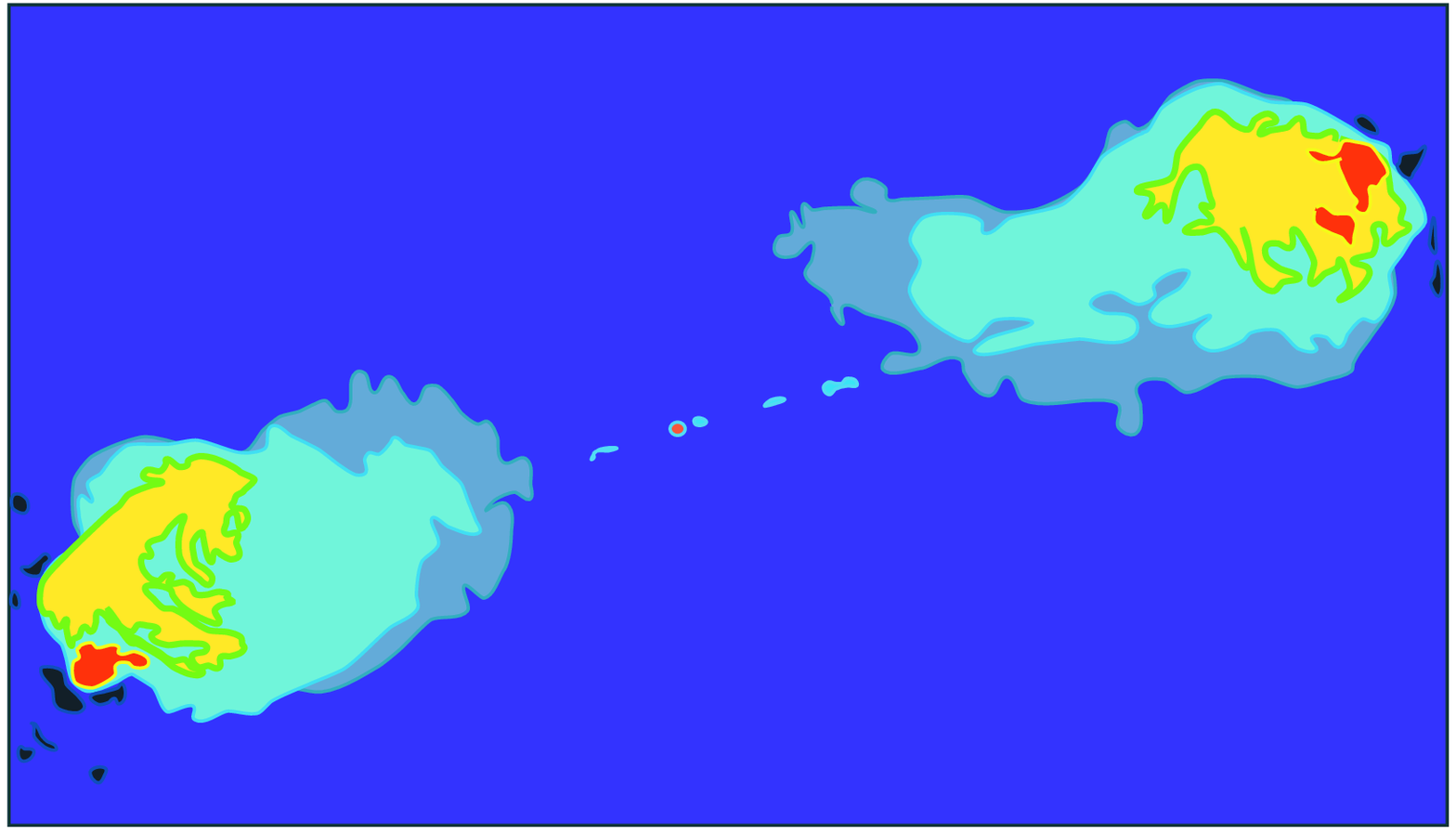}{9}{cyga} {The radio jet and lobes in the hyperluminous
  radio galaxy \astrobj{Cygnus A}. The hotspots in the two lobes, the
  core region and the jets are clearly visible.  (Reproduced from an
  image courtesy of NRAO/AUI.)}

Here, we show that our perception of a hypothetical object crossing
our field of vision at a constant superluminal speed is remarkably
similar to a pair of symmetric hotspots departing from a fixed point
with a decelerating rate of angular separation.

\image{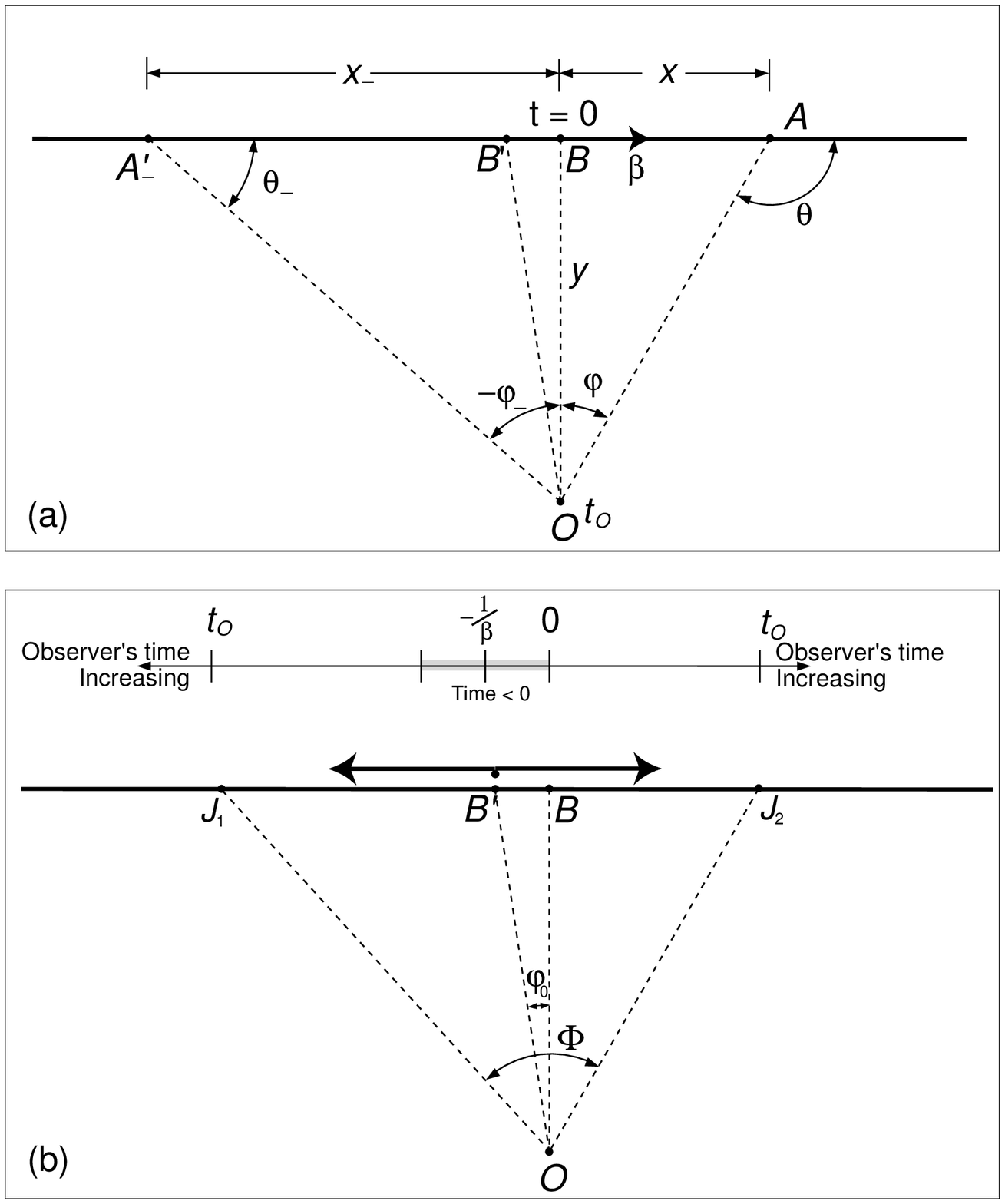}{9}{d} {The top panel (a) shows an object flying along
  $A_-'BA$ at a constant superluminal speed.  The observer is at $O$.
  The object crosses $B$ (the point of closest approach to $O$) at
  time $t=0$.  The bottom panel (b) shows how the object is perceived
  by the observer at $O$.  It first appears at $B'$, then splits into
  two.  The two apparent objects seem to go away from each other
  (along $J_1$ and $J_2$) as shown. This perceptual effect is best
  illustrated using an animation, which can be found at
  {\tt http://TheUnrealUniverse.com/anim.shtml}.}

Consider an object moving at a superluminal speed as shown in
\figref{d}(a).  The point of closest approach is $B$.  At that point,
the object is at a distance of $y$ from the observer at $O$.  Since
the speed is superluminal, the light emitted by the object at some
point $B'$ (before the point of closest approach $B$) reaches the
observer {\em before\/} the light emitted at $A_-'$.  This reversal
creates an illusion of the object moving in the direction from $B'$ to
$A_-'$, while in reality it is moving in the opposite direction from
$A_-'$ to $B'$.  This effect is better illustrated using an
animation.\footnote{The perceptual effect of a superluminal object
  appearing as two objects is much easier to illustrate using an
  animation, which can be found at the author's web site:
  {\tt http://TheUnrealUniverse.com/anim.html.}}

We use the variable $\ta$ to denote the observer's time.  Note that,
by definition, the origin in the observer's time axis is set when the
object appears at $B$.  $\phi$ is the observed angle with respect to
the point of closest approach $B$. $\phi$ is defined as $\theta -
\pi/2$ where $\theta$ is the angle between the object's velocity and
the observer's line of sight.  $\phi$ is negative for negative time
$t$.

Appendix~\ref{sec:super} readily derives a relation between $\ta$ and
$\phi$.
\begin{equation}
 \ta = y\left( \frac{\tan\phi}\beta + \frac{1}{\cos\phi} - 1\right)
 \label{eqn.3}
\end{equation}
Here, we have chosen units such that $c = 1$, so that $y$ is also the
time light takes to traverse $BO$.  The origin of the observer's time
is set when the observer sees the object at $B$.  \ie $\ta = 0$ when
the light from the point of closest approach $B$ reaches the observer.

The actual plot of $\phi$ as a function of the observer's time is
given in \figref{PhiVsTp0} for different speeds $\beta$.  Note that
for subluminal speeds, there is only one angular position for any
given $\ta$.  For subluminal objects, the observed angular position
changes almost linearly with the observed time, while for superluminal
objects, the change is parabolic.  The time axis scales with $y$.

\Eqnref{eqn.3} can be approximated using a Taylor series
expansion as:
\begin{equation}
\ta \approx y\left(\frac\phi\beta +
  \frac{\phi^2}{2}\right)\label{eqn.4}
\end{equation}
From the quadratic \eqnref{eqn.4}, one can easily see that the minimum
value of $\ta$ is $\ta_\text{min} = -y/2\beta^2$ and it occurs at
$\phi_{0}=-1/\beta$.  Thus, to the observer, the object first appears
(as though out of nowhere) at the position $\phi_0$ at time
$\ta_\text{min}$.  Then it appears to stretch and split, rapidly at
first, and slowing down later.

The angular separation between the objects flying away from each other
is:
\begin{equation}
\nonumber
 \Phi = \frac{2}{\beta}\sqrt{1+\frac{2\beta^2}{y}\ta} =
 \frac{2}{\beta}\left(1+\beta\phi\right)
\end{equation}
And the rate at which the separation occurs is:
\begin{equation}
\nonumber
 \frac{d\Phi}{d\ta} = \sqrt{\frac{2}{y t_\text{age}}} =
 \frac{2\beta}{y\left(1+\beta\phi\right)} \label{eqn.5}
\end{equation}
where $ t_\text{age} = \ta - \ta_\text{min}$, the apparent age of the
symmetric object. (The derivations of these equations can be found in
Appendix~\ref{sec:super}.)

\image{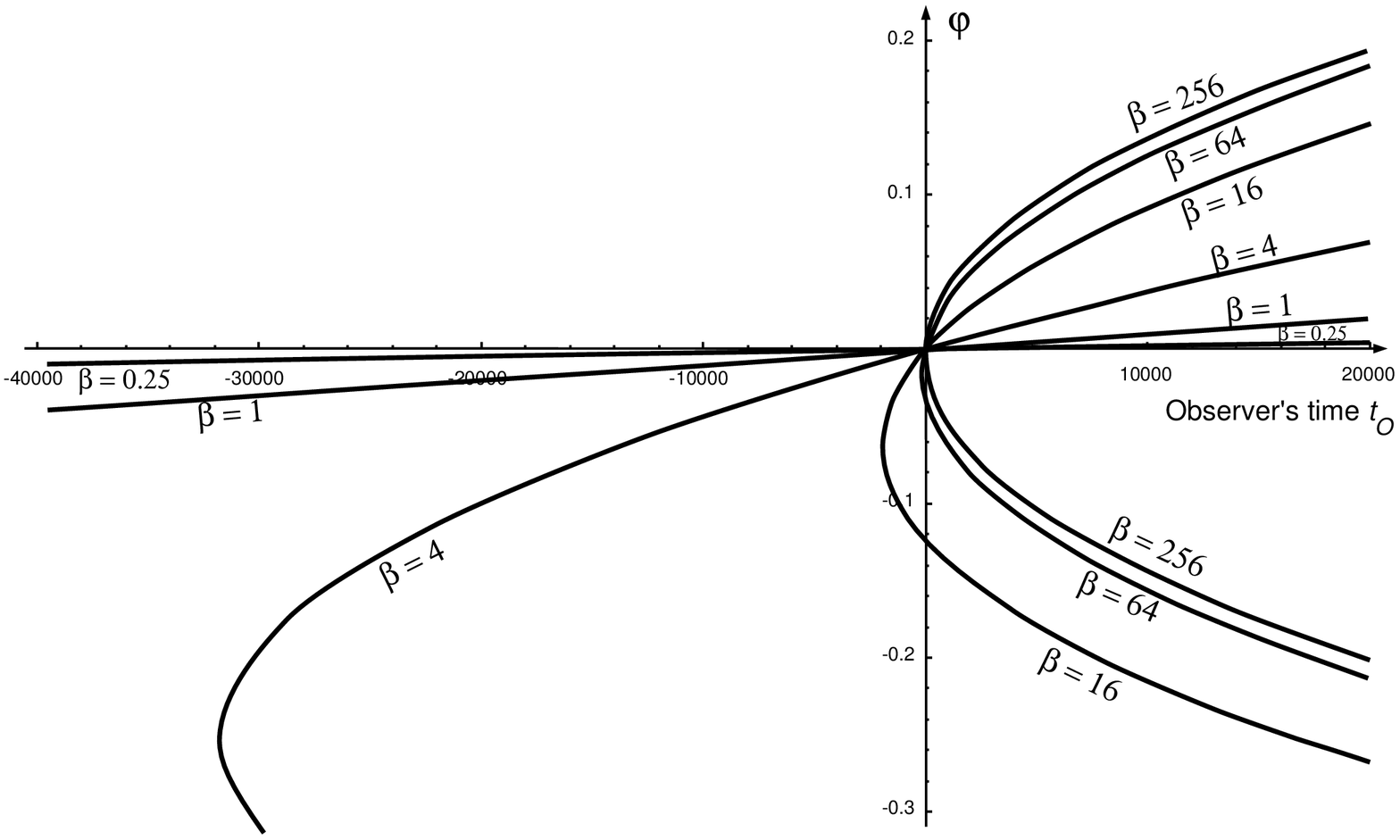}{9}{PhiVsTp0} {The apparent angular positions of an
  object traveling at different speeds at a distance $y$ of one
  million light years from us.  The angular positions ($\phi$ in
  radians) are plotted against the observer's time $\ta$ in years.}

This discussion shows that a single object moving across our field of
vision at superluminal speed creates an illusion of an object
appearing at a certain point in time, stretching and splitting into
two and then moving away from each other.  This time evolution of the
two objects is given in \eqnref{eqn.3}, and illustrated in the bottom
panel of \figref{d}(b).  Note that the apparent time $\ta$ (as
perceived by the observer) is reversed with respect to the real time
$t$ in the region $A_-$ to $B'$.  An event that happens near $B'$
appears to happen before an event near $A_-$.  Thus, the observer may
see an apparent violation of causality, but it is only a part of the
light travel time effect.

If there are multiple objects, moving as a group, at roughly constant
superluminal speed along the same direction, they will appear as a
series of objects materializing at the same angular position and
moving away from each other sequentially, one after another.  The
apparent knot in one of the jets always has a corresponding knot in
the other jet. In fact, the appearance of a superluminal knot in one
of the jets with no counterpart in the opposite jet, or a clear
movement in the angular position of the ``core'' (at point $B'$) will
invalidate our model.

\section{Redshifts of the Hotspots}
In the previous section, we showed how a hypothetical superluminal
object appears as two objects receding from a core.  Now we consider
the time evolution of the redshift of the two apparent objects (or
hotspots).  Since the relativistic Doppler shift equation is not
appropriate for our considerations (because we are working with
hypothetical superluminal objects), we need to work out the
relationship between the redshift ($\z$) and the speed ($\beta$) as we
would do for sound.  This calculation is done in Appendix~\ref{sec:z}:
\begin{eqnarray}
\nonumber
1 \,+\, \z \,&=&\, \left|1 \,-\, \beta\cos\theta\right| \\
\nonumber
\,&=&\, \left|1  \,+\, \beta\sin\phi\right|\\
\,&=&\, \left|1 \,+\, \frac{\beta^2t}{\sqrt{\beta^2t^2 + y^2}}\right|
\label{eqn.z0}
\end{eqnarray}

We can explain the radio frequency spectra of the hotspots as
extremely redshifted black body radiation because $\beta$ can be
enormous in our model of extragalactic radio sources.  Note that the
limiting value of $|1+\z|$ is approximately equal to $\beta$, which
gives an indication of the speeds required to push the black body
radiation of a typical star to the RF region. Since the speeds
($\beta$) involved are typically extremely large, and we can
approximate the redshift as:
\begin{equation}
\nonumber
1 \,+\, \z \,\approx\,  \left|\beta\phi\right| \,\approx\,
\frac{\left|\beta\Phi\right|}{2}
\end{equation}
Assuming the object to be a black body similar to the sun, we can
predict the peak wavelength (defined as the wavelength at which the
luminosity is a maximum) of the hotspots as:
\begin{equation}
\nonumber
\lambda_\text{max} \,\approx\, (1+\z) 480nm \,\approx\,
\frac{\left|\beta\Phi\right|}{2} 480nm \label{eqn.z1}
\end{equation}
where $\Phi$ is the angular separation between the two hotspots.

This equation shows that the peak RF wavelength increases linearly
with the angular separation.  If multiple hotspots can be located in a
twin jet system, their peak wavelengths will depend only on their
angular separation, in a linear fashion.  Such a measurement of the
emission frequency as $\phi$ increases along the jet is clearly seen
in the photometry of the jet in \astrobj{3C 273}.\cite{RF2UV}
Furthermore, if the measurement is done at a single wavelength,
intensity variation can be expected as the hotspot moves along the
jet.  In other words, measurements at higher wavelengths will find the
peak intensities farther away from the core region, which is again
consistent with observations.

\section{Gamma Ray Bursts}

The evolution of redshift of the thermal spectrum of a hypothetical
superluminal object also holds the explanation for gamma ray bursts
(GRBs).

\image{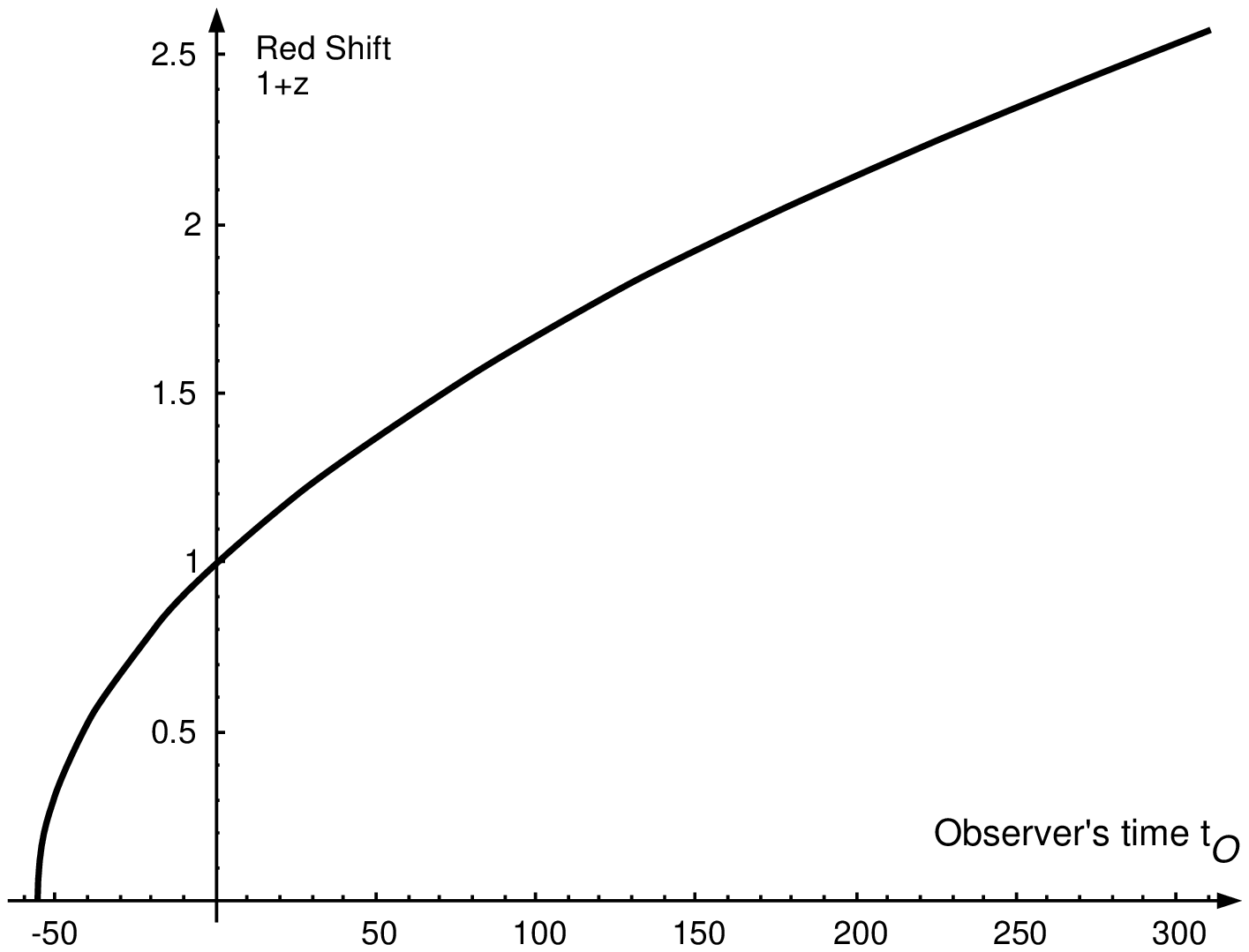}{8}{zVsTp}{Time evolution of the redshift from a
  superluminal object.  It shows the redshifts expected from an object
  moving at $\beta = 300$ at a distance of ten million light years
  from us.  The X axis is the observer's time in years.  (Since the X
  axis scales with time, it is also the redshift from an object at 116
  light days --ten million light seconds-- with the X axis
  representing $\ta$ in seconds.)}

The evolution of GRB can be made quantitative because we know the
dependence of the observer's time $\ta$ and the redshift $1+\z$ on the
real time $t$ (\eqnsref{eqn.3}{eqn.z0}).  From these two, we can
deduce the observed time evolution of the redshift (see
Appendix~\ref{sec:dz}).  We have plotted it parametrically in
\figref{zVsTp} that shows the variation of redshift as a function of
the observer's time ($\ta$).  The figure shows that the observed
spectra of a superluminal object is expected to start at the
observer's time $\ta_\text{min}$ with heavy (infinite) blue shift.
The spectrum of the object rapidly softens and soon evolves to zero
redshift and on to higher values.  The rate of softening depends on
the speed of the underlying superluminal object and its distance from
us.  The speed and the distance are the only two parameters that are
different between GRBs and symmetric radio sources in our model.

Note that the X axis in \figref{zVsTp} scales with time.  We have
plotted the variation of the redshift ($1+\z$) of an object with
$\beta = 300$ and $y =$ ten million light years, with X axis is $\ta$
in years.  It is also the variation of the redshift of an object at $y
=$ ten million light seconds (or 116 light days) with X axis in
seconds.  The former corresponds to symmetric jets and the latter to a
GRB.  Thus, for a GRB, the spectral evolution takes place at a much
faster pace.  Different combinations of $\beta$ and $y$ can be fitted
to describe different GRB spectral evolutions.

The observer sees no object before $\ta_\text{min}$.  In other words,
there is a definite point in the observer's time when the GRB is
``born'', with no indication of its impending birth before that time.
This birth does not correspond to any cataclysmic event (as would be
required in the collapsar/hypernova or the ``fireball'' model) at the
distant object.  It is nothing but an artifact of our perception.

In order to compare the time evolution of the GRB spectra to the ones
reported in the literature, we need to get an analytical expression
for the redshift ($\z$) as a function of the observer's time ($\ta$).
This can be done by eliminating $t$ from the equations for $\ta$ and
$1+\z$ (\eqnsref{eqn.3}{eqn.z0}), with some algebraic manipulations as
shown in Appendix~\ref{sec:dz}.  The algebra can be made more
manageable by defining $\tau = y/\beta$, a characteristic time scale
for the GRB (or the radio source).  This is the time the object would
take to reach us, if it were coming directly toward us.  We also
define the age of the GRB (or radio source) as $t_\text{age} = \ta -
\ta_\text{min}$.  This is simply the observer's time ($\ta$) shifted
by the time at which the object first appears to him
($\ta_\text{min}$).  With these notations (and for small values $t$),
it is possible to write the time dependence of $\z$ as:
\begin{equation}
  \label{eqn.z}
1 + \z = \left| {1 + \frac{{\beta ^2 \left( { - \tau  \pm \sqrt
            {2 \beta  t_{\text{age}}} } \right)}}{{\beta t_{\text{age}}
        + \tau /2 \mp \sqrt {2 \beta t_{\text{age}}}  + \beta ^2 \tau
      }}} \right|
\end{equation}
for small values of $t \ll \tau$.

Since the peak energy of the spectrum is inversely proportional to the
redshift, it can be written as:
\begin{equation}
  \label{eqn.Epk0}
  E_\text{pk}(t_{\text{age}}) = \frac{ E_\text{pk}(\ta_\text{min})}{1 + C_1
  \,\sqrt{\frac{t_{\text{age}}}{\tau}} + C_2\,\frac{t_{\text{age}}}{\tau}}
\end{equation}
where $C_1$ and $C_2$ are coefficients to be estimated by the Taylor
series expansion of \eqnref{eqn.z} or by fitting.

Ref.~\refcite{GRB6} have studied the evolution of the peak energy
($E_\text{pk}(t)$), and modeled it empirically as:
\begin{equation}
  \label{eqn.Epk}
  E_\text{pk}(t) = \frac{ E_\text{pk,0}}{(1+t/\tau)^\delta}
\end{equation}
where $t$ is the time elapsed after the onset ($= t_\text{age}$ in our
notation), $\tau$ is a time constant and $\delta$ is the hardness
intensity correlation (HIC).  Ref.~\refcite{GRB6} reported seven
fitted values of $\delta$.  We calculate their average as $\delta =
1.038\pm0.014$, with the individual values ranging from $0.4$ to
$1.1$.  Although it may not rule out or validate either model within
the statistics, the $\delta$ reported may fit better to
\eqnref{eqn.Epk0}.  Furthermore, it is not an easy fit because there
are too many unknowns.  However, the similarity between the shapes of
\eqnsref{eqn.Epk0}{eqn.Epk} is remarkable, and points to the agreement
between our model and the existing data.

\section{Conclusions}
In this article, we looked at the spatio-temporal evolution of a
supersonic object (both in its position and the sound frequency we
hear).  We showed that it closely resembles GRBs and DRAGNs if we were
to extend the calculations to light, although a luminal boom would
necessitate superluminal motion and is therefore forbidden.

This difficulty notwithstanding, we presented a unified model for
Gamma Ray Bursts and jet like radio sources based on bulk superluminal
motion.  We showed that a single superluminal object flying across our
field of vision would appear to us as the symmetric separation of two
objects from a fixed core. Using this fact as the model for symmetric
jets and GRBs, we explained their kinematic features quantitatively.
In particular, we showed that the angle of separation of the hotspots
was parabolic in time, and the redshifts of the two hotspots were
almost identical to each other. Even the fact that the spectra of the
hotspots are in the radio frequency region is explained by assuming
hyperluminal motion and the consequent redshift of the black body
radiation of a typical star.  The time evolution of the black body
radiation of a superluminal object is completely consistent with the
softening of the spectra observed in GRBs and radio sources. In
addition, our model explains why there is significant blue shift at
the core regions of radio sources, why radio sources seem to be
associated with optical galaxies and why GRBs appear at random points
with no advance indication of their impending appearance.

Although it does not address the energetics issues (the origin of
superluminality), our model presents an intriguing option based on how
we would perceive hypothetical superluminal motion.  We presented a
set of predictions and compared them to existing data from DRAGNs and
GRBs. The features such as the blueness of the core, symmetry of the
lobes, the transient $\gamma$ and X-Ray bursts, the measured evolution
of the spectra along the jet all find natural and simple explanations
in this model as perceptual effects.  Encouraged by this initial
success, we may accept our model based on luminal boom as a working
model for these astrophysical phenomena.

It has to be emphasized that perceptual effects can masquerade as
apparent violations of traditional physics.  An example of such an
effect is the apparent superluminal
motion,\cite{superluminal1,superluminal4,M87} which was explained and
anticipated within the context of the special theory of
relativity\cite{einstein} even before it was actually
observed.\cite{rees}  Although the observation of superluminal motion
was the starting point behind the work presented in this article, it is
by no means an indication of the validity of our model.  The
similarity between a sonic boom and a hypothetical luminal boom in
spatio-temporal and spectral evolution is presented here as a curious,
albeit probably unsound, foundation for our model.

One can, however, argue that the special theory of relativity (SR)
does not deal with superluminality and, therefore, superluminal motion
and luminal booms are not inconsistent with SR.  As evidenced by the
opening statements of Einstein's original paper,\cite{einstein} the
primary motivation for SR is a covariant formulation of Maxwell's
equations, which requires a coordinate transformation derived based
partly on light travel time (LTT) effects, and partly on the
assumption that light travels at the same speed with respect to all
inertial frames.  Despite this dependence on LTT, the LTT effects are
currently assumed to apply on a space-time that obeys SR.  SR is a
redefinition of space and time (or, more generally, reality) in order
to accommodate its two basic postulates. It may be that there is a
deeper structure to space-time, of which SR is only our perception,
filtered through the LTT effects. By treating them as an optical
illusion to be applied on a space-time that obeys SR, we may be double
counting them. We may avoid the double counting by disentangling the
covariance of Maxwell's equations from the coordinate transformations
part of SR. Treating the LTT effects separately (without attributing
their consequences to the basic nature of space and time), we can
accommodate superluminality and obtain elegant explanations of the
astrophysical phenomena described in this article. Our unified
explanation for GRBs and symmetric radio sources, therefore, has
implications as far reaching as our basic understanding of the nature
of space and time.

\appendix

\section{Mathematical Details}
\subsection{Doppler Shift}
\label{sec:z}

\image{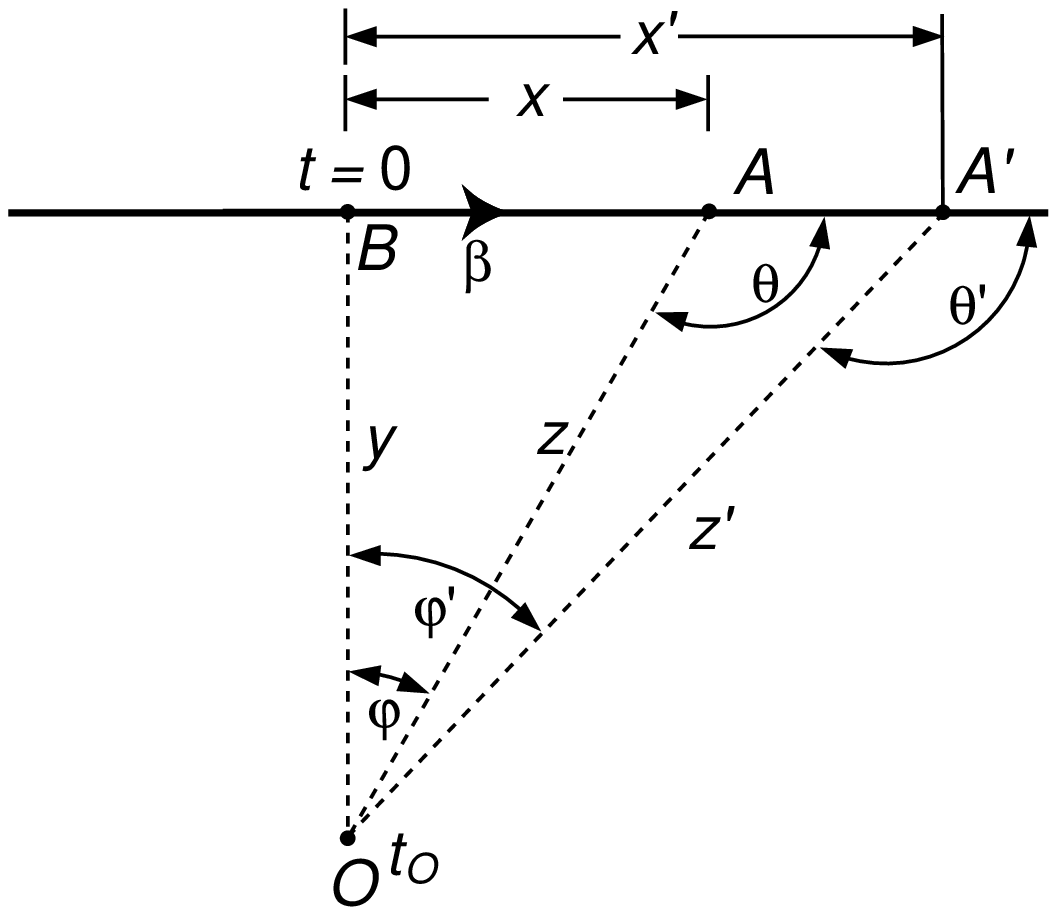}{7}{c} {The object is flying along $BAA'$, the observer
  is at $O$.  The object crosses $B$ (the point of closest approach) at
  time $t=0$.  It reaches $A$ at time $t$.  A photon emitted at $A$
  reaches $O$ at time $\ta$, and a photon emitted at $A'$ reaches $O$
  at time $\ta'$.}

We refer to \figref{c} and start by defining the real speed of the
object as:
\begin{equation}
\label{eqn.b}
v \,=\, \beta\,c \,=\, \frac{x'\,-\,x}{t'-t}
\end{equation}
But the speed it {\em appears\/} to have will depend on when the
observer senses the object at $A$ and $A'$.  The apparent speed of the
object is:
\begin{equation}
\label{eqn.bp}
v' \,=\, \bp\,c \,=\, \frac{x' \,-\, x}{\ta' \,-\,\ta}
\end{equation}
We also have
\begin{eqnarray}
\nonumber
\ta \,&=&\, t+\frac{z}{c}\\
\nonumber
 \ta'\,&=&\, t' + \frac{z'}{c}\\
\Rightarrow \ta'-\ta \,&=&\, t' - t + \frac{z'-z}{c}
\end{eqnarray}
Thus,
\begin{eqnarray}
\nonumber
\frac{\beta}{\bp} \,&=&\, \frac{\ta'-\ta}{t'-t} \\
\nonumber
\,&=&\, 1 + \frac{z'-z}{c(t'-t)} \\
\nonumber
\,&=&\, 1 - \frac{x-x'}{c(t'-t)}\cos\theta \\
\,&=&\, 1 - \beta\cos\theta
\end{eqnarray}
which gives,
\begin{eqnarray}
\nonumber
\bp &\,=\,&\frac{\beta}{1\,-\,\beta \cos\theta}\\
\beta &\,=\,&\frac{\bp}{1\,+\,\bp \cos\theta}
\end{eqnarray}
and,
\begin{eqnarray}
\nonumber
 \frac{\bp}{\beta}  &\,=\,& \frac{1}{1-\beta\,\cos\theta} \\
\nonumber
&\,=\,& 1+\bp\,\cos\theta\\
&\,=\,& \sqrt{\frac{1+\bp\,\cos\theta}{1-\beta\,\cos\theta}}
\label{eqn.7}\end{eqnarray}

Redshift ($\z$) defined as:
\begin{equation}
1 \,+\, \z \,=\, \frac{\lp}{\lambda}
\end{equation}
where $\lp$ is the measured wavelength and $\lambda$ is the known
wavelength.  In \figref{c}, the number of wave cycles created in
time $t'-t$ between $A$ and $A'$ is the same as the number of wave
cycles sensed at $O$ between $\ta'$ and $\ta$.  Substituting the
values, we get:
\begin{equation}
\frac{(t'-t)\, c}{\lambda} \,=\, {\frac{(\ta'\,-\,\ta)\,c}
{\lp}}
\end{equation}
Using the definitions of the real and apparent speeds
from \eqnsref{eqn.b}{eqn.bp}, it is easy to get:
\begin{equation}
\frac{\lp}{\lambda} \,=\, \frac{\beta}{\bp}
\end{equation}
Using the relationship between the real speed $\beta$ and the
apparent speed $\bp$ from \eqnref{eqn.7},
we get:
\begin{eqnarray}
\nonumber
1 \,+\, \z \,&=&\, \frac{1}{1 \,+\, \bp\cos\theta}\\
 \,&=&\, 1 \,-\,\beta\cos\theta
\end{eqnarray}
As expected, $\z$ depends on the longitudinal component of the
velocity of the object.  Since we allow superluminal speeds in this
calculation, we need to generalize this equation for $\z$ noting that
the ratio of wavelengths is positive.  Taking this into account, we
get:
\begin{eqnarray}
\nonumber
1 \,+\, \z \,&=&\, \left|\frac{1}{1 \,+\,
      \bp\cos\theta}\right| \\
\,&=&\, \left|1 \,-\, \beta\cos\theta\right| \label{eqn.8}
\end{eqnarray}
For a receding object $\theta=\pi$.  If we consider only subluminal
speeds, we can rewrite this as:
\begin{eqnarray}
\nonumber
1 \,+\, \z \,&=&\, \frac{1}{1 \,-\, \bp} \\
\nonumber
\,&=&\, 1 \,+\, \beta\\
\nonumber
(1 \,+\, \z)^2 \,&=&\, \frac{1 \,+\, \beta}{1 \,-\, \bp}
\end{eqnarray}
Or,
\begin{equation}
1 \,+\, \z \,=\, \sqrt{\frac{1+\beta}{1 \,-\,
      \bp}}
\end{equation}
If we were to mistakenly assume that the speed we observe is the real
speed, then this becomes the relativistic Doppler formula:
\begin{equation}
1 \,+\, \z \,=\, \sqrt{\frac{1+\beta}{1 \,-\,
      \beta}}
\end{equation}
\subsection{Kinematics of Superluminal Objects}
\label{sec:super}
\image{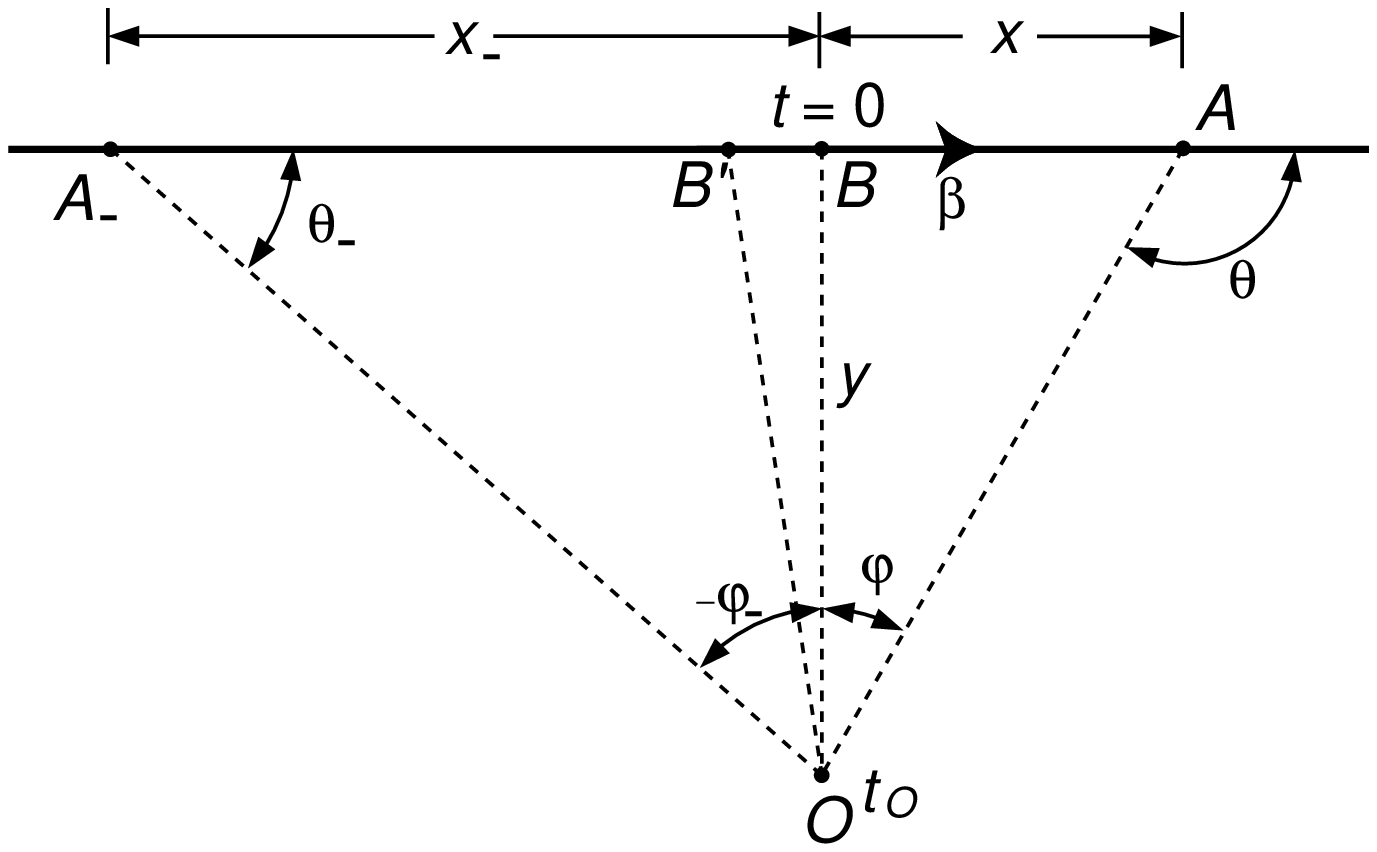}{8}{dd} {An object flying along $A_-BA$ at a constant
  superluminal speed.  The observer is at $O$.  The object crosses $B$
  (the point of closest approach to $O$) at time $t=0$.}

The derivation of the kinematics is based on \figref{dd}.  Here, an
object is moving at a superluminal speed along $A_-BA$. At the point
of closest approach, $B$, the object is a distance of $y$ from the
observer at $O$. Since the speed is superluminal, the light emitted by
the object at some point $B'$ (before the point of closest approach
$B$) reaches the observer {\em before\/} the light emitted at $A_-$.
This gives an illusion of the object moving in the direction from $B'$
to $A_-$, while in reality it is moving from $A_-$ to $B'$.

Observed angle $\phi$ is measured with respect to the point of closest
approach $B$ and is defined as $\theta - \pi/2$ where $\theta$ is the
angle between the object's velocity and the observer's line of sight.
$\phi$ is negative for negative time $t$.  We choose units such that
$c = 1$ for simplicity and denote the observer's time by $\ta$.  Note
that, by definition, the origin in the observer's time, $\ta$ is set
to the instant when the object appears at $B$.

The real position of the object at any time $t$ is:
\begin{equation}
 x = y\tan\phi = \beta t
\end{equation}
Or,
\begin{equation}
 t = \frac{y\tan\phi}{\beta}
\end{equation}
A photon emitted by the object at $A$ (at time $t$) will reach $O$
after traversing the hypotenuse.  A photon emitted at $B$ will reach
the observer at $t = y$, since we have chosen $c = 1$.  Since we define
the observer's time $\ta$ such that the time of arrival is $t = \ta +
y$, then we have:
\begin{equation}
 \ta = t + \frac{y}{\cos\phi} - y
\end{equation}
which gives the relation between $\ta$ and $\phi$.
\begin{equation}
 \ta = y\left( \frac{\tan\phi}\beta + \frac{1}{\cos\phi} - 1\right)
\end{equation}
Expanding the equation for $\ta$ to second order, we get:
\begin{equation}
 \ta = y\left(\frac\phi\beta + \frac{\phi^2}{2}\right)\label{eqn.9}
\end{equation}
The minimum value of $\ta$ occurs at $\phi_{0}=-1/\beta$ and it is
$\ta_\text{min} = -y/2\beta^2$.  To the observer, the object first
appears at the position $\phi=-1/\beta$.  Then it appears to stretch
and split, rapidly at first, and slowing down later.

The quadratic \eqnref{eqn.9} can be recast as:
\begin{equation}
  \label{eqn.q1}
  1+\frac{2\beta^2}{y}\ta = \left(1+\beta\phi\right)^2
\end{equation}
which will be more useful later in the derivation.

The angular separation between the objects flying away from each other
is the difference between the roots of the quadratic
\eqnref{eqn.9}:
\begin{eqnarray}
\nonumber
 \Phi \,&=&\, \phi_1-\phi_2 \\
\nonumber
\,&=&\, \frac{2}{\beta}\sqrt{1+\frac{2\beta^2}{y}\ta} \\
\,&=&\, \frac{2}{\beta}\left(1+\beta\phi\right)
\end{eqnarray}
making use of \eqnref{eqn.q1}.  Thus, we have the angular separation
either in terms of the observer's time ($\Phi(\ta)$) or the angular
position of the object ($\Phi(\phi)$) as illustrated in
Figure~\ref{phiphi}.

\image{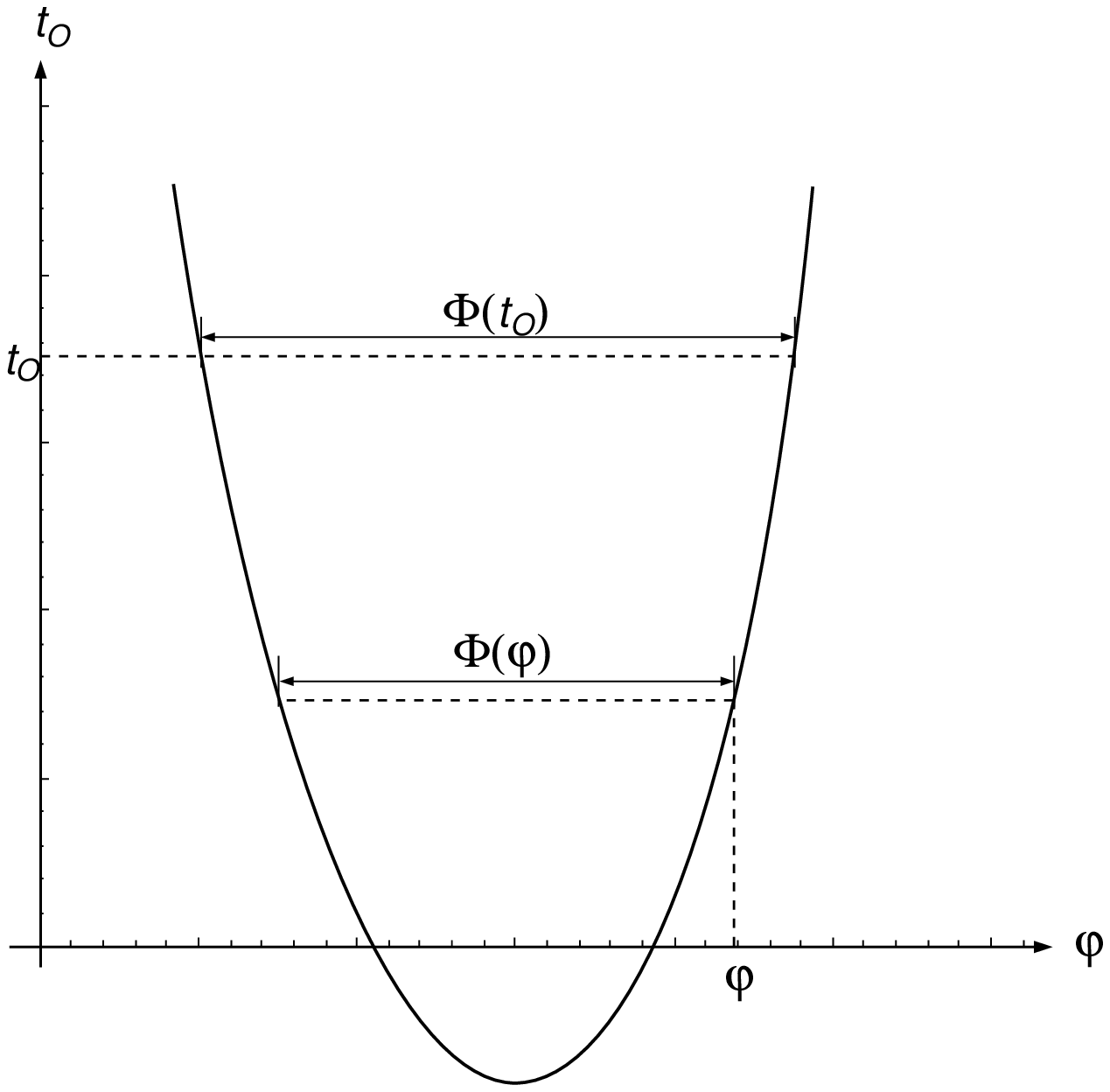}{8}{phiphi} {Illustration of how the angular separation
  is expressed either in terms of the observer's time ($\Phi(\ta)$) or
  the angular position of the object ($\Phi(\phi)$)}

The rate at which the angular separation occurs is:
\begin{eqnarray}
\nonumber
 \frac{d\Phi}{d\ta} \,&=&\, \frac{2\beta}{y\sqrt{1+\frac{2\beta^2}{y}\ta}}
 \\
\,&=&\, \frac{2\beta}{y\left(1+\beta\phi\right)}
\end{eqnarray}
Again, making use of \eqnref{eqn.q1}.  Defining the apparent age of
the radio source $ t_\text{age} = \ta - \ta_\text{min}$ and knowing
$\ta_\text{min} = -y/2\beta^2$, we can write:
\begin{eqnarray}
\nonumber
 \frac{d\Phi}{d\ta} \,&=&\, \frac{2\beta}{y\sqrt{1+\frac{2\beta^2}{y}\ta}}\\
\nonumber
\,&=&\, \frac{2\beta}{y\sqrt{1-\frac{\ta}{\ta_\text{min}}}}\\
\nonumber
\,&=&\, \sqrt{\frac{4\beta^2}{y^2}\,\times\,\frac{-\ta_\text{min}}{\ta-\ta_\text{min}}}\\
\,&=&\,\sqrt{\frac{2}{y\, t_\text{age}}}
\end{eqnarray}

\subsection{Time Evolution of the Redshift}
\label{sec:dz}
As shown before in \eqnref{eqn.8}, the redshift $\z$ depends
on the real speed $\beta$ as:
\begin{equation}
1 \,+\, \z \,=\, \left|1 \,-\, \beta\cos\theta\right| \,=\, \left|1
  \,+\, \beta\sin\phi\right|\label{eqn.10}
\end{equation}
For any given time ($\ta$) for the observer, there are two solutions
for $\phi$ and $\z$.  $\phi_1$ and $\phi_2$ lie on either side of
$\phi_0 = 1/\beta$.  For $\sin\phi > -1/\beta$, we get
\begin{equation}
 1+\z_2 = 1+\beta\sin\phi_1
\end{equation} and for $\sin\phi < -1/\beta$,
\begin{equation}
 1+\z_1 = -1 - \beta\sin\phi_2
\end{equation}
Thus, we get the difference in the redshift between the two hotspots
at $\phi_1$ and $\phi_2$ as:
\begin{equation}
 \Delta\z \approx 2 + \beta(\phi_1+\phi_2)
\end{equation}
We also have the mean of the solutions of the quadratic ($\phi_1$ and
$\phi_2$) equal to the position of the minimum ($\phi_0$):
\begin{equation}
\frac{\phi_1 + \phi_2}{2} = -\frac{1}{\beta}
\end{equation}
Thus $\phi_1+\phi_2 = -2/\beta$ and hence $\Delta\z = 0$.  The two
hotspots will have identical redshifts, if terms of $\phi^3$ and above
are ignored.

As shown before (see \eqnref{eqn.10}), the redshift $\z$
depends on the real speed $\beta$ as:
\begin{equation}
1 \,+\, \z \,=\,\left|1 \,+\, \beta\sin\phi\right| \,=\, \left|1 \,+\,
  \frac{\beta^2t}{\sqrt{\beta^2t^2 + y^2}}\right|
\end{equation}
Since we know $\z$ and $\ta$ functions of $t$, we can plot their
inter-dependence parametrically.  This is shown in \figref{zVsTp} of
the article.

It is also possible to eliminate $t$ and derive the dependence of
$1+\z$ on the apparent age of the object under consideration,
$t_\text{age} = \ta - t_\text{min}$.  In order to do this, we first
define a time constant $\tau = y/\beta$.  This is the time the object
would take to reach us, if it were flying directly toward us.  Keeping
in mind that the new variable is related to $t_\text{age}$ through $
\ta_\text{min} = - y/2\beta ^2 = - \tau /\beta$, let's get an
expression for $t/\tau$:
\begin{eqnarray}
\nonumber
 \ta \,&=&\, t + \sqrt {\beta ^2 t^2  + y^2 }  - y \\
\nonumber
  \,&=&\, t + \beta \tau \sqrt {1 + \frac{{t^2 }}{{\tau ^2 }}}  - \beta \tau  \\
\nonumber
  \,&\approx&\, t + \frac{{\beta t^2 }}{{2\tau }} \\
  \Rightarrow \frac{t}{\tau } \,&=&\, \frac{{ - 1 \pm \sqrt {1 +
  \frac{{2\beta t_\text{age} }}{\tau }} }}{\beta }
\label{eqn.ttau}
\end{eqnarray}
Note that this is valid only for $t \ll \tau$.  Now we collect the
terms in $t/\tau$ in the equation for $1+\z$:
\begin{eqnarray}
\nonumber
 \ta \,&=&\, t + \sqrt {\beta ^2 t^2  + y^2 }  - y \\
\nonumber
 \Rightarrow \sqrt {\beta ^2 t^2  + y^2 }  \,&=&\, \ta - t + y \\
\nonumber
 1 + z \,&=&\, \left| {1 + \frac{{\beta ^2 t}}{{\sqrt {\beta ^2 t^2  + y^2 } }}} \right| \\
\nonumber
  \,&=&\, \left| {1 + \frac{{\beta ^2 t}}{{\ta - t + y}}} \right| \\
  \,&=&\, \left| {1 + \frac{{\beta ^2 \frac{t}{\tau
 }}}{{\frac{{t_{\text{age}} }}{\tau } - \frac{1}{{2\beta }} - \frac{t}{\tau } + \beta }}}
 \right|
\label{eqn.zz}
\end{eqnarray}
As expected, the time variables always appear as ratios like $t/\tau$,
giving confidence that our choice of the characteristic time scale is
probably right.  Finally, we can substitute $t/\tau$ from
\eqnref{eqn.ttau} in \eqnref{eqn.zz} to obtain:
\begin{equation}
1 + \z = \left| {1 + \frac{{\beta ^2 \left( { - \tau  \pm \sqrt
            {2 \beta  t_{\text{age}}} } \right)}}{{\beta t_{\text{age}}
        + \tau /2 \mp \sqrt {2 \beta t_{\text{age}}}  + \beta ^2 \tau
      }}} \right|
\end{equation}


\end{document}